\documentclass[published]{JHEP}
\JHEP{08(1999)010}
\usepackage{epsfig,amssymb}

\newcommand{\apj}[3]{\emph{Astrophys.\ J.}\ {\bf #1} (#2) #3}

\def\lsim{\lesssim}
\def\gsim{\gtrsim}

\title{Implications of a new solar system population
of neutralinos on indirect detection rates}

\author{\leavevmode\parbox{.7\textwidth}{Lars Bergstr{\"o}m$^a$, Thibault Damour$^b$, 
	Joakim Edsj{\"o}$^a$, Lawrence M.\ Krauss$^c$ and Piero
	Ullio$^a$}\\
	\leavevmode\llap{$^a$}Dept.\ of Physics, Stockholm University, Box 6730,
	SE-113~85~Stockholm, Sweden\\[1mm]
	\leavevmode\llap{$^b$}Institut des Hautes Etudes
	Scientifiques,\\ 
	35 route de Chartres 91440 Bures sur Yvette, France\\[1mm]
	\leavevmode\llap{$^c$}Departments of Physics and Astronomy, 
	Case Western Reserve University\\	
	10900 Euclid Ave, Cleveland OH 44106-7079\\[1mm]
	E-mail: \email{lbe@physto.se}, \email{damour@ihes.fr},
	\email{edsjo@physto.se}, \email{krauss@theory1.phys.cwru.edu}, 
	\email{piero@physto.se}}

\abstract{Recently, a new Solar System population of weakly
interacting massive particle (WIMP) dark matter has been proposed to
exist.  We investigate the implications of this population on indirect
signals in neutrino telescopes (due to WIMP annihilations in the
Earth) for the case when the WIMP is the lightest neutralino of the
MSSM, the minimal supersymmetric extension of the standard model.  The
velocity distribution and capture rate of this new population is
evaluated and the flux of neutrino-induced muons from the center of
the Earth in neutrino telescopes is calculated.  The strength of the
signal is very sensitive to the velocity distribution of the new
population.  We analytically estimate this distribution using the
approximate conservation of the component of the WIMP angular momentum
orthogonal to the ecliptic plane.  The non-linear problem of combining
a fixed capture rate from the standard galactic WIMP population with
one rising linearly with time from the new population to obtain the
present-day annihilation rate in the Earth is also solved
analytically.  We show that the effects of the new population can be
crucial for masses below around 150 GeV, where enhancements of the
predicted muon flux from the center of the Earth by up to a factor of
100 compared to previously published estimates occur.  As a result of
the new WIMP population, the next generation of neutrino telescopes
should be able to probe a much larger region of parameter space in the
mass range 60-130 GeV.}

\received{may 24, 1999}
\accepted{August 19, 1999}

\begin{document}

\section{Introduction} 

It has been known for almost 15 years that Weakly Interacting Massive
Particles (WIMPs) can elastically scatter inside the Sun and Earth,
leading to their subsequent capture and annihilation in the cores of
these objects, producing an indirect neutrino signature that might be
accessible to neutrino telescopes~\cite{genrefs}.  Recently, however,
it has been demonstrated that the scattering process in the Sun can
populate orbits which subsequently result in a bound Solar System
population of WIMPs~\cite{dk1,dk2} and which can be comparable in
spectral density, in the region of the Earth, to the Galactic halo
WIMP population.  This new population consists of WIMPs that have
scattered in the outer layers of the Sun and due to perturbations by
the other planets (mainly Jupiter) evolve into bound orbits which do
not cross the Sun but do cross the Earth's orbit.  This population of
WIMPs should have a completely different velocity distribution than
halo WIMPs and will thus have quite different capture probabilities in
the Earth.  The predicted WIMP abundance, and spectrum, relevant for
direct detection have been calculated~\cite{dk1,dk2}.  Here we focus
on capture in the Earth, and the predicted indirect neutrino
signature.  Other studies of solar system populations of WIMPs can be
found in \cite{genrefs,otherpop,Gould91}.

Following~\cite{dk1,dk2}, we consider here a special WIMP candidate,
the neutralino, which arises naturally in supersymmetric extensions of
the standard model.  We evaluate the capture of neutralinos and the
\pagebreak[3]
resulting neutrino-induced muon flux within the Minimal Supersymmetric
extension of the Standard Model (MSSM) (for a review of neutralino
dark matter, see ref.~\cite{jkg}).  We find that this flux is
sensitively dependent upon the Solar System WIMP velocity
distribution, and can exceed by two orders of magnitude that predicted
for Galactic halo WIMPs. Although our numerical results apply only to
supersymmetric WIMPs, we expect our qualitative conclusions to be
generally valid for any WIMP model of dark matter.

\section{Definition of the supersymmetric model}
\label{sec:MSSMdef}

We work in the Minimal Supersymmetric Standard Model (MSSM)\@.  In
general, the MSSM has many free parameters, but with customary
assumptions~\cite{jkg} (which do not significantly affect the general
results of this paper) we can reduce the number of parameters to
seven: the Higgsino mass parameter $\mu$, the gaugino mass parameter
$M_{2}$, the ratio of the Higgs vacuum expectation values $\tan
\beta$, the mass of the $CP$-odd Higgs boson $m_{A}$, the scalar mass
parameter $m_{0}$ and the trilinear soft SUSY-breaking parameters
$A_{b}$ and $A_{t}$ for the third generation.  In particular, we do
not impose any restrictions from supergravity other than gaugino mass
unification.  We have made some scans in parameter space without the
GUT relation $3M_1=5M_2\tan^2\theta_W$ for the gaugino mass parameters
$M_1$ and $M_2$\@.  This mainly has the effect of allowing lower
neutralino masses to escape the LEP bounds and is not very important
for this study.  Hence, the GUT relation for $M_1$ and $M_2$ is kept
throughout this paper.  For a more detailed definition of the
parameters and a full set of Feynman rules, see
refs.~\cite{coann,jephd}. Of course we assume $R$-parity conservation,
which ensures that the lightest supersymmetric particle is stable and
therefore an excellent dark matter candidate. Its relic abundance
today is fixed by its freeze-out abundance in the early Universe,
which in turn is given by the interplay between the interaction rate
and the expansion rate at a temperature around $T\sim m_\chi/20$,
where $m_\chi$ is the neutralino mass (i.e.\ it is non-relativistic at
freeze-out and would act as cold dark matter in structure formation).

The lightest (and, with $R$-parity conservation, stable)
supersymmetric particle is in most models the lightest neutralino,
which is a superposition of the superpartners of the gauge and Higgs
fields,
\begin{equation}
  \tilde{\chi}^0_1 =
  N_{11} \tilde{B} + N_{12} \tilde{W}^3 +
  N_{13} \tilde{H}^0_1 + N_{14} \tilde{H}^0_2\,.
\end{equation}
It is convenient to define the gaugino fraction of the lightest neutralino,
\begin{equation}
  Z_g = |N_{11}|^2 + |N_{12}|^2\,.
\end{equation}
For the masses of the neutralinos and charginos we use the one-loop
corrections as given in ref.~\cite{neuloop} and for the Higgs boson
masses we use the leading log two-loop radiative corrections,
calculated within the effective potential approach given in
ref.~\cite{carena}.

\TABLE[t]{%
  \begin{tabular}{|r|rrrrrrr|} \hline
  Parameter & $\mu$ & $M_{2}$ &
  $\tan \beta$ & $m_{A}$ & $m_{0}$ & $A_{b}/m_{0}$ & $A_{t}/m_{0}$ \\
  Unit & GeV & GeV & 1 & GeV & GeV & 1 & 1 \\ \hline
  Min & -50000 & -50000 & 1.0  & 0     & 100   & -3 & -3 \\
  Max & 50000  & 50000  & 60.0 & 10000 & 30000 &  3 &  3 \\  \hline
  \end{tabular}%
\caption{The ranges of parameter values used in our scans of the MSSM
parameter space. Note that several special scans aimed at interesting
regions of the parameter space have been performed.  In total we have
generated approximately $1.4\times 10^5$ models which are not excluded by
accelerator searches.}  \label{tab:scans}}

We have made extensive scans of the model parameter space, some
general and some specialized to interesting regions, where high rates
are possible.  In total we have made 33 different scans of the
parameter space.  The scans were done randomly and were mostly
distributed logarithmically in the mass parameters and in $\tan
\beta$\@.  For some scans the logarithmic scan in $\mu$ and $M_{2}$
was replaced by a logarithmic scan in the more physical parameters
$m_{\chi}$ and $Z_{g}/(1-Z_{g})$, where $m_{\chi}$ is the neutralino
mass. Combining all the scans, the overall limits of the seven MSSM
parameters we use are given in table~\ref{tab:scans}.

We check each model to see if it is excluded by the most recent
accelerator constraints, of which the most important ones are the LEP
bounds~\cite{lepbounds} on the lightest chargino mass,
\begin{equation}
  m_{\chi_{1}^{+}} > \left\{ \begin{array}{lcl}
  91 {\rm ~GeV} & \,, \qquad & | m_{\chi_{1}^{+}} -
  m_{\chi^{0}_{1}} |
  > 4 {\rm ~GeV} \\
  85 {\rm ~GeV} & \,, \qquad & {\rm otherwise}
  \end{array} \right.
\end{equation}
and on the lightest Higgs boson mass $m_{H_{2}^{0}}$ (which range from
72.2--88.0 GeV depending on $\sin (\beta-\alpha)$ with $\alpha$ being
the Higgs mixing angle) and the constrains from $b \to s \gamma$
\cite{cleo}.  For each allowed model we calculate the relic density of
neutralinos $\Omega_\chi h^2$, where $\Omega_\chi$ is the density in
units of the critical density and $h$ is the present Hubble constant
in units of $100$ km s$^{-1}$ Mpc$^{-1}$\@.  We use the formalism of
ref.~\cite{GondoloGelmini} for resonant annihilations, threshold
effects, and finite widths of unstable particles and we include all
two-body tree-level annihilation channels of neutralinos.  We also
include so-called coannihilation processes in the relic density
calculation according to the analysis of Edsj\"o and Gondolo~\cite{coann}.

Present observations favor $h=0.65 \pm 0.1$, and a total matter density
$\Omega_{M}=0.3\pm 0.1$, of which baryons may contribute 0.02 to 0.08
\cite{cosmparams}.
Not to be overly restrictive, we accept
$\Omega_\chi h^2$ in the range from $0.025$ to $1$ as cosmologically
interesting.  The lower bound is somewhat arbitrary as there may be
several different components of non-baryonic dark matter, but we
demand that neutralinos are at least as abundant as required to make
up the dark halos of galaxies.  In principle, neutralinos with
$\Omega_\chi h^2<0.025$ would still be relic particles, but only
making up a small fraction of the dark matter of the Universe,
so we do not consider this alternative here.

\section{Speed distribution of the new neutralino population}

The distribution of Solar System WIMPs relevant to terrestrial capture
arises from scattering of Galactic halo WIMPs into highly elliptical
Solar System orbits, with semi-major axes in the range $ a_{\rm earth} < a
< 2.6 a_{\rm earth}$.  These are then perturbed by Jupiter and the other
planets into orbits which do not intersect the Sun.
Analytical methods were used in refs.~\cite{dk1,dk2} to estimate the basic
features of this distribution. Here we shall relax an approximation
(exactly radial orbits) made in refs.~\cite{dk1,dk2}
and derive an improved analytical estimate of this distribution.
We nevertheless expect, because of their
origin in Solar scattering, that most of these
orbits will be close to
radial in the vicinity of the Earth.

In ref.~\cite{Gould91}, the phase space distribution of WIMPs at the Earth
was investigated including the effects of the Earth being deep in the
Sun's potential well. There it was found that even though the velocity
distribution of WIMPs at the Earth is different than it would be in
free space, both Jupiter, the Earth and Venus will disturb the orbits
of these unbound WIMPs into bound orbits that have the same phase
space distribution as would be the case in free space. This relies on
the fact that the time-scales for transferring WIMPs between these
unbound and bound orbits are shorter than the age of the Earth so that
an equilibrium occurs.  This is valid for low velocities relative
to the Earth only  (which
are the most important ones for capture) and it was found that for
some higher-velocity orbits (velocities $\gsim$ 27 km/s),
the time-scale for filling or depleting these bound
orbits is too long. It turns out that the (nearly) radial orbits
of interest for this work
happen to be in precisely this regime.  More precisely, our new
population is distributed along a thin vertical ``needle'', located
within the ``unfilled bound orbits'' region of fig.~3 of
ref.~\cite{Gould91}, starting at $u = 30$ km/s on the left part of the
horizontal axis.  This is of course an advantage for detection since
it means that these WIMPs will stay in these orbits without being much
affected by diffusion within the Solar System.

The detection rates we discuss below are very sensitive to the actual
form of the velocity distribution, especially at low velocities (with
respect to the Earth). The reason for this is easy to understand from
simple kinematics. A very heavy WIMP (heavier than iron) which
encounters the Earth cannot be stopped by a single encounter if its
velocity is large.  (And the optical depth for repeated scatterings is
very small.) Therefore, it is the strength of the distribution at low
energies which determines the capture rate and therefore the muon
neutrino rate from the center of the Earth.  An exception to the rule
of only low velocities being important occurs if the WIMP has a mass
which closely matches one of the most abundant elements of the
Earth. Then kinematics allows all the kinetic energy of the WIMP to be
transferred to a nucleus in a single collision.  As we shall discuss
in detail below, conservation of the z component of the angular
momentum constrains very much the component of the velocity of the WIMPs along
the motion of the Earth,
and thereby the lowest attainable speed with respect to the Earth.  As
a consequence, we shall see that only neutralinos of the new
population lighter than around 150 GeV will be captured by the Earth.

Let us now turn to the best estimate we can presently make of the
velocity distribution, as seen in an Earth frame, of the new
neutralino population. In refs.~\cite{dk1,dk2}, as a first
approximation, a model was used where the WIMP orbits were nearly
radial (i.e.\ with a fixed eccentricity, very near $e=1$).  Here, we
shall go beyond this approximation by taking into account the
distribution in the $z$ component of the angular momentum (which will
turn out to be the crucial one for our purpose). Let us first define
our notation.

In an heliocentric frame, the WIMP phase-space distribution function
would read $dN = f ({\bf x} , {\bf v}) \, d^3 {\bf x} \wedge d^3 {\bf
v}$, where ${\bf x}$ denotes the WIMP heliocentric position and ${\bf
v}$ its heliocentric velocity.  We could start from $f({\bf x} , {\bf
v})$, written as a function of the integrals of motion of the WIMP (so
that it satisfies Liouville's equation), to derive the capture rate in
the Earth. It is, however, more convenient to take advantage from the
start of the simplification brought by the spherical symmetry of the
Earth. Let ${\bf u} = {\bf v} - {\bf v}_E$ denote the incoming
velocity (far from the Earth) of the WIMP in an Earth reference frame
(here ${\bf v}_E$ denotes the heliocentric velocity of the Earth). The
WIMP capture rate by the spherically symmetric Earth depends only on
the angular average of the velocity distribution, and therefore is a
function only of the distribution of ``speeds'' $u \equiv \vert {\bf
u} \vert$. We can write the speed distribution of the local WIMP
number density as
\begin{equation}
  dn (u) = n(a_1) \, d \, \widehat{n} (u) \, , \label{eq3.1}
\end{equation}
where $n(a_1)$, with $a_1 \equiv 1AU$ denoting the radius of the
Earth's orbit, is the total space density around the Earth (WIMPs per
cm$^3$), and where $d \, \widehat{n} (u)$ is the \emph{normalized}
speed distribution: $\int d \, \widehat{n} (u) = 1$. The integrated
WIMP space density $n(a_1)$ has been estimated in
refs.~\cite{dk1,dk2}. We shall quote the result below when we need
it. We focus first on the normalized speed distribution $d \,
\widehat{n} (u)$.

At any moment we can introduce an heliocentric spatial reference frame
$(x,y,z)$ such that the $x$-axis is in the Sun-Earth radial direction, and the
$y$-axis is directed along the orbital motion of the Earth. In this frame
\begin{equation}
u^2 = ({\bf v} - {\bf v}_E)^2 = {\bf v}^2 + v_E^2 - 2 \, v_E \, v_y \, .
\label{eq3.2}
\end{equation}
As we are considering a fixed value $r \equiv \vert {\bf x} \vert =
a_1$ of the WIMP radial position, the squared heliocentric velocity
${\bf v}^2$ can be expressed in terms of the semi-major axis of the
WIMP orbit:
\begin{equation}
{\bf v}^2 (r = a_1) = v_E^2 \left( 2 - \frac{a_1}{a} \right)  . \label{eq3.3}
\end{equation}
On the other hand, the longitudinal (``along the track'') velocity
$v_y$ entering eq.~(\ref{eq3.2}) can be expressed in terms of the
$z$-component $J_z$ of the angular momentum of the WIMP, $v_y = J_z /
a_1$ (note that the $z$ axis is orthogonal to the ecliptic plane).
Finally, the (square of the) local speed of the WIMP can be entirely
expressed in terms of the two adiabatic invariants $a$ and $J_z$
(i.e.\ the two Delaunay action variables $L = \sqrt{G_N \, M_{\rm Sun}
\, a}$ and $H = \sqrt{G_N \, M_{\rm Sun} \, a (1-e^2)} \cos i$ in the
notation of ref.~\cite{dk2}):
\begin{equation}
u^2 = v_E^2 \left( 3 - \frac{a_1}{a} - \frac{2 \, J_z}{a_1 \, v_E} \right)  .
\label{eq3.4}
\end{equation}
We recall that $v_E = (G_N \, M_{\rm Sun} / a_1)^{1/2} = 27.98$~km~s$^{-1}$
denotes  the Earth orbital speed, and $a_1 = r_E = 1\ AU$.

Let us recall from ref.~\cite{dk2} that, to a good approximation, the
two WIMP action variables $L$ and $H$ are secularly conserved (modulo
some small random-walk-type diffusion caused by near encounters with
the inner planets), while the third action variable $G = \sqrt{G_N \,
M_{\rm Sun} \, a (1-e^2)}$ will secularly evolve under the influence
of planetary perturbations. [The averaged planetary Hamiltonian
$\overline{\cal H}_p$ (Kozai Hamiltonian) was discussed in section~3
of~\cite{dk2}.] The evolution of $G$ means that the eccentricity of
the WIMP orbit oscillates between values very near one and smaller
values. We see from eq.~(\ref{eq3.4}) that these eccentricity
oscillations have no effect on the local speed distribution. The
latter depends only on the well conserved quantities $L$ and $H$, or
$a$ and $J_z$. We can then estimate the present speed distribution by
means of the initial distribution of $a$ and $J_z$. [``Initial'' means
here at the time when the considered WIMP of the new population exited
from the Sun after having been captured by scattering on a nucleus in
the outskirts of the Sun.] The initial $z$-component of the WIMP
angular momentum was
\begin{equation}
J_z^0 = \sqrt{G_N \, M_{\rm Sun} \, a_0 (1-e_0) (1+e_0)} \cos i_0 \, .
\label{eq3.5}
\end{equation}
Here, $a_0 (1-e_0)$ is the initial perihelion distance which is approximately
equal to the (effective) Sun radius $\overline{R}_S = 0.907 \, R_{\rm Sun}$.
The remaining factor $1+e_0$ can be approximated by 2, so that
\begin{equation}
J_z^0 \simeq \sqrt{2 G_N \, M_{\rm Sun} \, \overline{R}_S} \cos i_0 \, ,
\label{eq3.6}
\end{equation}
where the cosine of the initial inclination can be (approximately) considered
as being a random variable uniformly distributed between $-1$ and $+1$.

Because of the overall (approximate) axial symmetry of the Solar
System, we expect the conservation law of $J_z$ to be rather accurate:
$J_z^{\rm now} \simeq J_z^0$. However, as the nonconservation of $J_z$
would have a crucial effect on the capture rate of WIMPs by the Earth
(by allowing for lower speeds in an Earth frame, through the last term
in eq.~(\ref{eq3.4})) we shall introduce a phenomenological parameter
$\lambda$ to parametrize a possible nonconservation of $J_z$. More
precisely, we shall assume that
\begin{equation}
J_z^{\rm now} = \lambda \, J_z^0 \, , \label{eq3.7}
\end{equation}
and study not only the ``standard'' case where $\lambda = 1$, but also the case
where, say, $\lambda = 2$, which corresponds to a 100\% violation of the
conservation of $J_z$. Then, eq.~(\ref{eq3.4}) reads
\begin{equation}
u^2 = v_E^2 \left( 3 - \frac{a_1}{a} - \varepsilon \, \cos i_0 \right) ,
\label{eq3.8}
\end{equation}
where
\begin{equation}
\varepsilon \equiv 2 \lambda \ \sqrt{ \frac{2 \, \overline{R}_S}{a_1}} \simeq
0.18377 \, \lambda \, . \label{eq3.9}
\end{equation}

Eq.~(\ref{eq3.8}) tells us that the distribution function of the
squared speed $u^2$ is obtained by taking the convolution between the
distribution of the variable $b \equiv a_1 / a$ (which is essentially
the WIMP heliocentric energy) and the (uniform) distribution of the
variable $c \equiv \varepsilon \, \cos i_0$. If we denote $b' \equiv b
+ \varepsilon \, \cos i_0$, the distribution $\varphi' (b') \, db'$ of
$b'$ is given in terms of the distribution $\varphi (b) \, db$ of $b
\equiv a_1 / a$ by
\begin{equation}
\varphi' (b') = \frac{1}{2 \, \varepsilon} \int_{-\varepsilon}^{\varepsilon}
\varphi (b'-\xi) \, d\xi \, . \label{eq3.10}
\end{equation}

The normalized distribution $\varphi (b) \, db$ (with $\int \varphi
(b) \, db =1$) has been derived in ref.~\cite{dk2}. It is written in
eq.~(6.7) there, in terms of the related variable $x = 2a / a_1 =
2/b$. (Ref.~\cite{dk2} made the approximation of nearly radial orbits
in deriving eq.~(6.7), but this approximation is not crucial for our
present purpose. The crucial new feature for the present work is the
allowance for the along-the-track term proportional to $J_z$ in
eq.~(\ref{eq3.4}) which is the only one which can significantly affect
neutralino capture by the Earth.) Transcribing eq.~(6.7) of~\cite{dk2}
in terms of the variable $b = 2/x$ yields
\begin{equation}
\varphi (b) \, db = N_b \ \frac{db}{b^{0.6} \, (2-b)^{1/2}} \ \theta \left( b -
\frac{1}{2.6} \right) \, \theta (2-b) \, , \label{eq3.11}
\end{equation}
with the normalization constant
\begin{equation}
N_b = \frac{2^{0.1}}{I_n (5.2)} = 0.456579 \, . \label{eq3.12}
\end{equation}
Here, as above, $\theta (x)$ denotes the step function ($\theta (x) = 1$ if
$x>0$, $\theta (x) = 0$ if $x<0$).

Eq.~(\ref{eq3.10}) then gives the distribution in $b'$, or equivalently in the
scaled squared speed
\begin{equation}
U \equiv \widehat{u}^2 \equiv \left( \frac{u}{v_E} \right)^2 = 3 - b' = 3 - b -
\varepsilon \, \cos i_0 \, . \label{eq3.13}
\end{equation}

Before the smearing of eq.~(\ref{eq3.10}) (due to the convolution with
the $J_z$-distribution) the original distribution $\varphi (b) \, db$
of eq.~(\ref{eq3.11}) featured both a cut-off at $b_1 = a_1 / a = 2$
and a cut-off at $b_2 = 1/2.6$. The former cut-off corresponds to a
lower-velocity cut-off at $u = v_E (3 - b_1)^{1/2} = v_E =
29.78$~km~s$^{-1}$. (It corresponds to radial orbits which barely
reach up to the Earth orbit.) In our present refined treatment, where
we take into account the along-the-track motion linked to the
distribution in $J_z$, this lower-velocity cut-off will be shifted to
a lower value (coming from a positive $J_z$ in eq.~(\ref{eq3.4})). For
capture by the Earth (which has a small escape velocity, ranging
between about 11 and 15~km~s$^{-1}$) the low-velocity part of the
speed distribution is the only one to play a crucial role. We can take
advantage of this fact to derive a simplified analytical expression of
the speed distribution $d \, \widehat n (U) = \varphi' (3-U) \, dU =
\varphi' (b') \, db'$. Indeed, we can approximate the smearing,
eq.~(\ref{eq3.10}), as acting only on the crucial low-velocity cut-off
factor $\theta (2-b) / (2-b)^{1/2}$ in eq.~(\ref{eq3.11}). The
smearing of the smoothly varying factor $1/b^{0.6}$ would only
introduce a small fractional correction of order $\varepsilon^2 \sim
3$\%. With this approximation the smearing integral of
eq.~(\ref{eq3.10}) can be explicitly computed. The final result for
the speed distribution reads
\begin{equation}
  d \, \widehat n = N'_{\varepsilon} \, f_{\varepsilon} (U) \, dU =
  N'_{\varepsilon}
  \, f_{\varepsilon} (\widehat{u}^2) \, 2 \, \widehat u \, d \, \widehat u\,,
  \label{eq3.14}
\end{equation}
where we recall that $U \equiv \widehat{u}^2 \equiv (u/v_E)^2$, where
$N'_{\varepsilon}$ is a normalization constant (equal to $N_b$,
eq.~(\ref{eq3.12}), modulo ${\cal O} (\varepsilon^2)$ fractional
corrections\footnote{In our calculations we used $N'_{\varepsilon} =
N_b = 0.456579$. We verified numerically that this induces an error of
less than 2\% for $\lambda=1$.})  and where $f_{\varepsilon} (U)$ is
explicitly given by
\begin{equation}
  f_{\varepsilon} (U) = \frac{\theta \, (U_{\rm max} - U)}{(3-U)^{0.6}} \
  \frac{1}{\varepsilon} \left[ \sqrt{U - 1 + \varepsilon} \ \theta \, (U - 1 +
  \varepsilon) - \sqrt{U - 1 - \varepsilon} \ \theta \, (U - 1 - \varepsilon)
  \right] \, . \label{eq3.15}
\end{equation}
Here $U_{\rm max} = 3 - 1/2.6 \simeq 2.6154$ is the upper cut-off, and
$\varepsilon$ is given by eq.~(\ref{eq3.9}) (with $\lambda = 1$ being
our standard estimate, and $\lambda = 2$ being an extreme
phenomenological possibility). Note that the speed distribution
extends only between a minimum cut-off at
\begin{equation}
  u_{\rm min} = v_E \ \sqrt{1-\varepsilon} \label{eq3.16}
\end{equation}
and a maximum at $u_{\rm max} = v_E \ \sqrt{U_{\rm max}} \simeq
48.16$~km~s$^{-1}$. (The exact maximum cut-off given by
eq.~(\ref{eq3.10}) is modified by ${\cal O} (\varepsilon)$ terms, but
this modification is unimportant for our present purpose.) The
numerical value of the lower speed cut-off, eq.~(\ref{eq3.16}), is
26.905~km~s$^{-1}$ for the standard case $\lambda = 1$, and
23.683~km~s$^{-1}$ for the extreme case $\lambda = 2$. The precise
value of this lower speed cut-off is important because it translates
into an \emph{upper} cut-off in the masses of the WIMPs that can be
captured by the Earth. Indeed, as will be recalled below, the
kinematics of the maximum energy loss in the collision of a WIMP of
mass $m_X$ with a nucleus of mass $m_A$ located at radius $r$ in the
Earth (with local escape velocity $v_{\rm esc} (r)$) is such that
capture is only possible if
\begin{equation}
  \beta_-^A \equiv \frac{4 \, m_X \, m_A}{(m_X - m_A)^2} \geq \left(
 \frac{u_{\rm min}}{v_{\rm esc} (r)} \right)^2 \, . \label{eq3.17}
\end{equation}
This implies the following upper limit on the mass of a WIMP
capturable by the Earth
\begin{equation}
m_X \leq \left( 1 + 2w + 2 \, \sqrt{w^2 + w} \right) m_A \, , \label{eq3.18}
\end{equation}
where we denoted $w \equiv (v_{\rm esc} (r) / u_{\rm min})^2$.  The
highest capturable mass is reached when the WIMP scatters on an iron
nucleus ($m_A = m_{\rm Fe} \simeq 52$~GeV)
located at the center of the Earth (where $v_{\rm esc} (0) \simeq
15$~km~s$^{-1}$).  In our standard case ($\lambda = 1$) where $u_{\rm
min} \simeq 26.905$~km~s$^{-1}$ this yields the upper limit
\begin{equation}
m_X \leq 2.90 \, m_{\rm Fe} \simeq 150 \, {\rm GeV} \, , \label{eq3.19}
\end{equation}
on the other hand in the case $\lambda = 2$ the limit is shifted to about
170 GeV.

\section{Overdensity of the new population}\label{s4}

We follow the notation of~\cite{dk1,dk2} and write the contribution
from this new population of WIMPs to the usual halo WIMP density as
\begin{equation}
  \delta_E \equiv \frac{n (a_1)}{n_X} \equiv \frac{\hbox{(secondary) WIMP
  density at the Earth}}{\hbox{halo WIMP density at infinity}} \, ,
  \label{eq5.11}
\end{equation}
where
\begin{equation}
  \delta_E = \frac{5.44 \times 10^{36}}{(v_o / 220 \, {\rm km\,s}^{-1})}
  \times
  g_{\rm tot} \, {\rm GeV} \, {\rm cm}^{-2} = \frac{0.212}{(v_o / 220 \,
  {\rm km\,s}^{-1})} \, g_{\rm tot}^{(-10)} \, . \label{eq5.15}
\end{equation}
Here, $g_{\rm tot}^{(-10)} \equiv 10^{10} \, g_{\rm tot} ({\rm
GeV})^3$, and $g_{\rm tot} = { \sum_A} \, (f_A / m_A) \,
\sigma_A \, K_A^s$, where $f_A$ is the mass fraction of element $A$ in
the Sun, and $K_A^s$ is the surface value of the capture function on
the element of mass number $A$ in the Sun according to eq.~(2.20) in
ref.~\cite{dk2}.

The scattering rate of WIMPs in the outer layers of the Sun (which
causes the fast halo WIMPs to lose enough energy to enter bound orbits
close to the Earth's orbit) is proportional to $\sigma_A K_A^s$, which
can be calculated once the parameters of the SUSY WIMP in question are
fixed. (For the elemental abundances in the Sun, we use the compilation of
Bahcall and Pinsonneault~\cite{jnb}.)

\FIGURE[t]{%
\epsfig{file=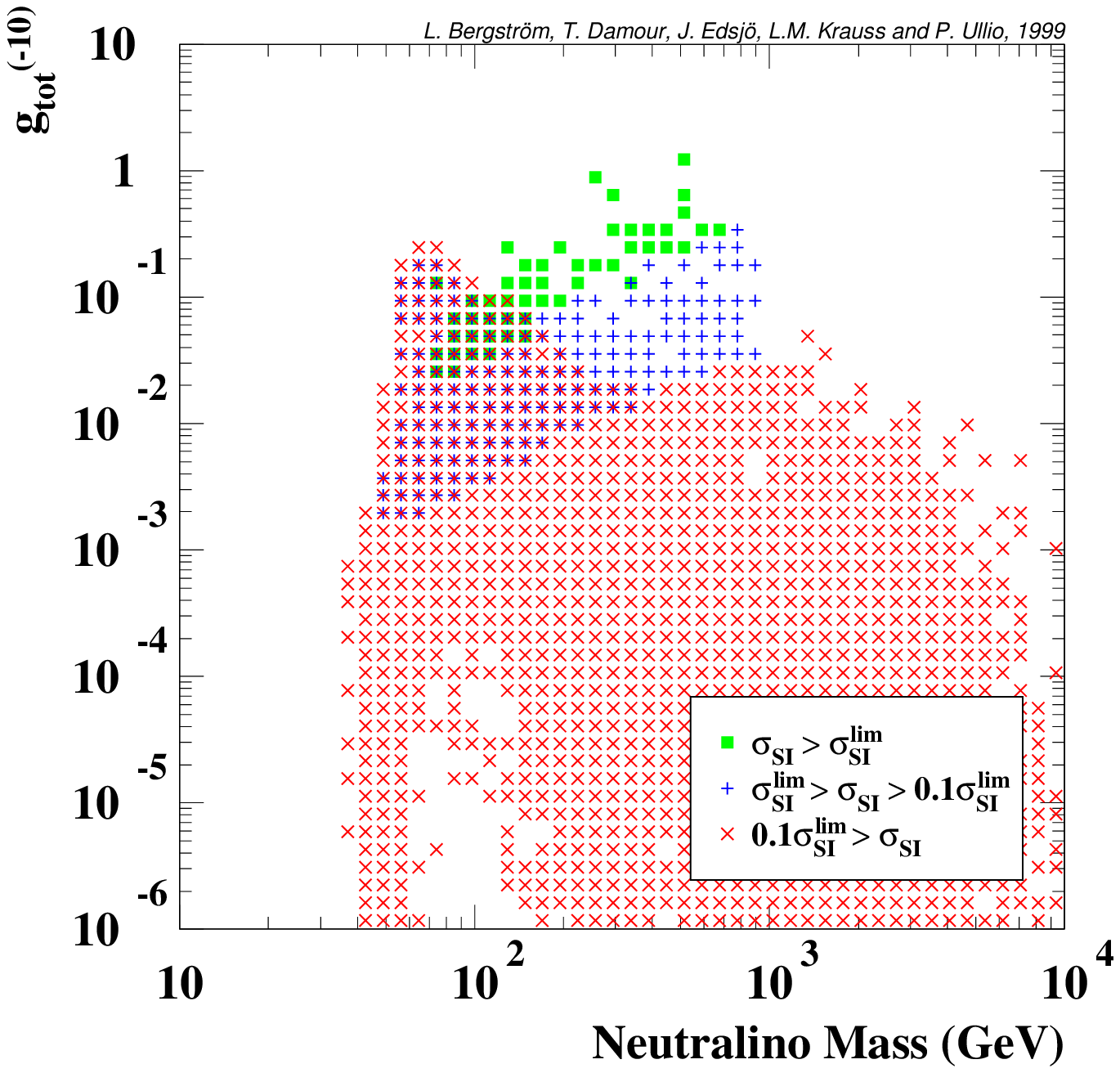,width=0.75\textwidth}
\caption{The value of the parameter $g_{{\rm tot}}^{(-10)}$, which is
related to the scattering probability near the Solar surface, as a
function of neutralino mass. The different symbols represent different
ranges for the spin-independent cross section, where $\sigma_{SI}^{\rm
lim}$ represents the most stringent claimed upper limit on this cross
section for an assumed halo density of 0.3 GeV/cm$^3$, given in
\protect\cite{uppbnd} (see text for a discussion).\label{fig:fig1}}}

In fig.~\ref{fig:fig1} we display the values of $g_{\rm tot}^{(-10)}$
versus neutralino mass for our set of supersymmetric models. As can be
seen, in some cases values approaching or even exceeding unity can be
obtained (confirming the results of~\cite{dk1,dk2}). The spread is
very large, however, and some models give orders of magnitude smaller
values. As would be expected, the models with the highest values of
$g_{\rm tot}^{(-10)}$ are the same models which give high scattering
rates in direct detection experiments
\pagebreak[3]
(the value of the local halo density is assumed to be 0.3~GeV~cm$^{-3}$).
This can be seen by the coding of the displayed points in terms of
one claimed limit~\cite{uppbnd} on the spin-independent scattering cross
section\footnote{Note that in ref.~\cite{dk1,dk2}
larger values of $g_{\rm tot}$ were obtained
because the possibility of smaller halo WIMP densities was allowed, and
also because more conservative claimed limits were used.  Here instead, in
order to be conservative in our estimate of predicted muon fluxes in
detectors, we use as our fiducial value of the WIMP scattering cross
section the most stringent claimed cross section upper limit~\cite{uppbnd}.}.

Note that the highest values of $g_{\rm tot}^{(-10)}$ allowed by this
claimed limit on the scattering rates is $g_{\rm
 tot~max}^{(-10)} \simeq 0.4$.  Such values imply only a rather modest
(but not negligible, and important as independent check) effect of the
new population in direct experiments (see refs.~\cite{dk1,dk2}). On
the other hand, we shall see below that they can imply a large effect
on indirect detection rates.

\section{Capture rate of WIMPs by the Earth}

The total space density of WIMPs, per speed element, in the vicinity of the
Earth reads
\begin{equation}
  \begin{array}[b]{rcl} dn_X^{\rm tot} = dn_X^{\rm old} + dn_X^{\rm new}
  & = & n_X \, d \, \widehat{n}^{\rm old} (u) + n^{\rm new} (a_1) \, d
  \, \widehat{n}^{\rm new} (u) 
\\ & = & n_X \left[d \, \widehat{n}^{\rm
  old} (u) + \delta_E \, d \, \widehat{n}^{\rm new} (u)\right] .
  \end{array} \label{eq5.1}
\end{equation}
The ``old'' distribution $d \, \widehat{n}^{\rm old} (u)$ can, to a
good approximation~\cite{Gould91}, be considered to be simply the
incoming Galactic one, shifted to the frame of the moving Earth. The
``new'' (secondary) distribution $d \, \widehat{n}^{\rm new} (u)$ is
that defined by eq.~(\ref{eq3.14}) above. The present overdensity
ratio $\delta_E$ is that discussed in section~\ref{s4} above. Let us now
write down the capture rate of WIMPs by the Earth~\cite{KSW,Gould87}.
We shall utilize the result of Gould~\cite{Gould87}, using the notation
of ref.~\cite{dk2}. As said above, we work here
with the angle-averaged speed distribution, which reads, in the
notation of ref.~\cite{dk2}, $dn_X (u) = 4\pi \, u^2 \, du \,
\overline{f}_{\infty} (u)$. Using eqs.~(2.20)--(2.22) of ref.\
\cite{dk2}, considered for $J_{\rm min} = 0$, we get for the capture
rate on element $A$:
\begin{equation}
  d \dot{N}_A = \frac{1}{\beta_+^A} \ d^3 {\bf x} \, n_A (r) \,
  \sigma_A \ \frac{dn_X (u)}{u} \ \exp \left[ -\frac{m_X}{2Q_A} \ (u^2
  + \alpha) \right] \, \theta_{\alpha} \, d \alpha \, , \label{eq5.2}
\end{equation}
with
\begin{equation}
  \theta_{\alpha} \equiv \theta \left[ \beta_-^A \left( v_{\rm esc}^2 (r) -
  \frac{\alpha}{\beta_+^A} \right) - u^2 \right] . \label{eq5.3}
\end{equation}
Here
\begin{equation}
  \beta_{\pm}^A \equiv \frac{4 \, m_X \, m_A}{(m_X \pm m_A)^2} \, , \label{eq5.4}
\end{equation}
$\sigma_A$ would be the total scattering cross section on nucleus
$m_A$ in absence of the form factor $F_A^2 (Q) = \exp \, (-Q / Q_A)$,
with $Q = E_{\rm before} - E_{\rm after}$ denoting the energy loss and
$Q_A$ the standard value of the coherence energy~\cite{jkg}, recalled
in eq.~(2.28) of~\cite{dk2}, and $\alpha$ is linked to the semi-major
axis of the WIMP, after capture by the Earth, by
\begin{equation}
  \alpha \equiv \frac{G_N \, M_{\rm Earth}}{a} \, . \label{eq5.5}
\end{equation}
Note that we work here in a geocentric frame (with ${\bf u} = {\bf
v}_X^{\infty} - {\bf v}_E$). The differential capture rate,
eq.~(\ref{eq5.2}), must be integrated over $\alpha$ and the volume of
the Earth, and must be summed over the label $A$ denoting the various
elements in the Earth. The integral over $\alpha$ is limited by the
theta function $\theta_{\alpha}$, eq.~(\ref{eq5.3}), and by the fact
that $\alpha$ must be positive (which expresses the fact that the WIMP
lost enough energy by scattering on element $A$ to end up being bound
to the Earth). These two constraints restrict the range of variation
of $\alpha$ to $0 < \alpha < \alpha_m (r,u)$, with
\begin{equation}
  \alpha_m (r,u) = \beta_+^A \, v_{\rm esc}^2 (r) -
\frac{\beta_+^A}{\beta_-^A} \, u^2 \, , \label{eq5.6}
\end{equation}
and imply the presence of a remaining theta function restricting the
allowed values of $u$:
\begin{equation}
  \theta_u \equiv \theta \, (\beta_-^A \, v_{\rm esc}^2 (r) - u^2) \,.
\label{eq5.7}
\end{equation}
Finally, integration over $\alpha$ (followed by integration over $d^3 {\bf x}$
and $dn_{X}(u)$,
and summation over $A$) leads to a total capture rate
\begin{equation}
  \begin{array}{rcl}
  C = \sum_A \dot{N}_A &=
  &\displaystyle\sum_A \frac{2Q_A}{\beta_+^A \, m_X} \int d^3 {\bf x} \, n_A (r) \,
  \sigma_A \, \frac{dn_X (u)}{u}\times \\[2ex]
  & & {}\times\left[ \exp \left( -
  \frac{m_X}{2Q_A} \, u^2 \right) - \exp \left( - \frac{m_X}{2Q_A} \,
  \beta_+^A (v_{\rm esc}^2 (r) + u^2) \right) \right] \theta_u \, .
  \end{array} \label{eq5.8}
\end{equation}

\TABLE[!t]{%
\begin{tabular}{|l|llll|} \hline
  Region & $f_{\rm O}$ & $f_{\rm Mg}$ & $f_{\rm Si}$ & $f_{\rm Fe}$ \\ \hline
  $r<3488$ km           & 0.05 & 0    & 0    & 0.90 \\
  3488 km $<r<$ 6378 km & 0.44 & 0.23 & 0.21 & 0.059 \\ \hline
\end{tabular}%
\caption{The mass fractions, $f_{i}$ of various elements in the
Earth. Data from ref.~\protect\cite{EncBrit}.\label{tab:earthcomp}}}

\FIGURE{\epsfig{file=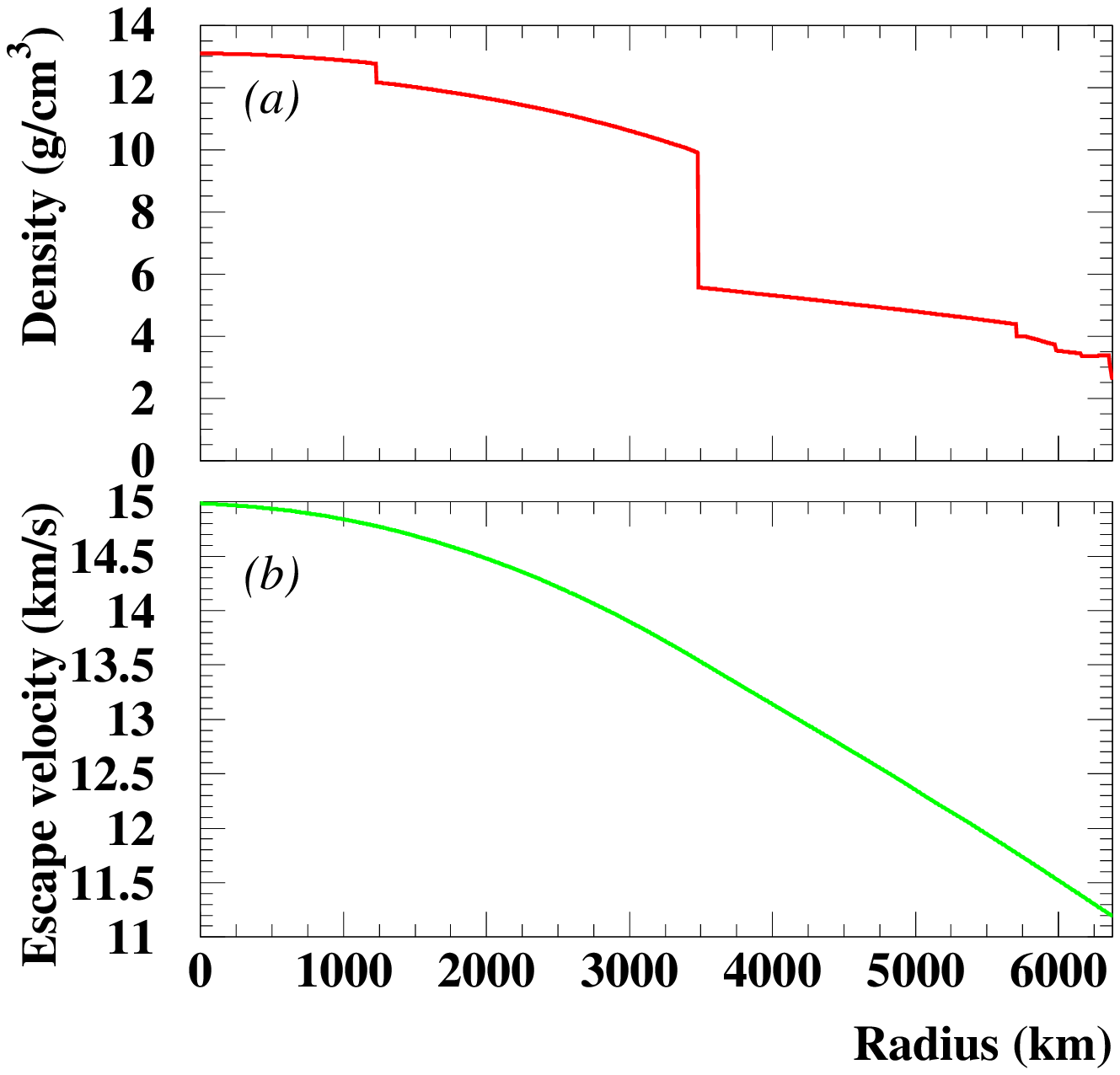,width=0.5\textwidth}%
\caption{Density (\emph{a}), and escape velocity (\emph{b}), as functions of the
  distance from the center of the Earth.\label{fig:densvesc}}}

This capture rate is a linear functional of the WIMP spectrum $dn_X
(u)$.  Therefore the decomposition, eq.~(\ref{eq5.1}), implies a
corresponding linear decomposition of the capture rate
\begin{equation}
  C^{\rm tot} = C^{\rm old} + C^{\rm new} \, . \label{eq5.9}
\end{equation}
To evaluate $C^{\rm new}$ we have to perform the integral in
eq.~(\ref{eq5.8}) explicitly.  Available analytical results in the lite\-ra\-tu\-re
do not apply since they assume a\linebreak Maxwell-Boltzmann velocity
dis\-tri\-bu-\linebreak tion.  Since we have spherical symmetry, we are left with a
double integral, one over the distance from the Earth's center, $r$,
and one over the velocity $u$.  For the composition and density of the
Earth as a function of $r$ we use the distributions given in
ref.~\cite{EncBrit}. From the density distribution, we can calculate
the gravitational potential as a function of $r$ and hence the escape
velocity $v_{\rm esc}(r)$. The density and escape velocity as a
function of $r$ are given in fig.~\ref{fig:densvesc}. In
table~\ref{tab:earthcomp} we show the mass fraction of the (for our
purpose) most important elements for different regions in the Earth.

\section{Neutralino annihilation rate in the Earth}

Let $N(t)$ be the total number of neutralinos trapped, at time $t$, in the core
of the Earth. The annihilation rate of neutralino pairs can be written as
\begin{equation}
\Gamma_a (t) = \frac{1}{2} \ C_a \, N^2 (t) \, . \label{eq6.1}
\end{equation}

The evolution of $N(t)$ is the result of the competition between capture and
annihilation:
\begin{equation}
\frac{dN}{dt} = C^{\rm tot} (t) - C_a \, N^2 \label{eq6.2}
\end{equation}
(the factor 2 difference in the annihilation terms in eqs.~(\ref{eq6.1}) and
(\ref{eq6.2}) comes from the fact that neutralinos annihilate in pairs). The
constant $C_a$ entering equations (\ref{eq6.1}) and (\ref{eq6.2}) is linked to
the annihilation cross-section $\sigma_a$, and to some effective volumes $V_j$,
$j=1,2$, taking into account the quasi-thermal distribution of neutralinos in
the Earth core:
\begin{equation}
C_a = \langle \sigma_a \, v \rangle \, \frac{V_2 }{ V_1^2} \, , \label{eq6.3}
\end{equation}
\begin{equation}
V_j \simeq 2.3 \times 10^{25} \, \left(\frac{j \, m_X}{ 10 \, {\rm GeV}}\right)^{-3/2} \, {\rm
cm}^3 \, . \label{eq6.4}
\end{equation}
Usually, the capture rate $C$ in eq.~(\ref{eq6.2}) is time-independent, being
given by the ``old'', Galactic WIMP population: $C^{\rm usual} = C^{\rm old} =
{\rm const}$. In that case, eq.~(\ref{eq6.2}) can be simply solved by
separating the two variables $N$ and $t$, i.e.
\begin{equation}
t = \int dt = \int \frac{d \, N}{C^{\rm old} - C_a \, N^2} \, . \label{eq6.5}
\end{equation}
This yields for the standard, ``old'' annihilation rate
\begin{equation}
\Gamma_a^{\rm old} = \frac{1}{2} \ C_a \, N^2 = \frac{1}{2} \ C^{\rm old} \,
\tanh^2 \, (\gamma \, t) \, , \label{eq6.6}
\end{equation}
with
\begin{equation}
\gamma \equiv (C^{\rm old} \, C_a)^{1/2} \, . \label{eq6.7}
\end{equation}
In the case considered here, where capture comes both from the direct
Galactic
population and from the secondary population discussed in refs.~\cite{dk1,dk2},
the total capture rate increases linearly with time, because of the
linear-in-time build up of the new population:
\begin{equation}
C^{\rm tot} (t) = C^{\rm old} + \dot{C} \, t \, , \label{eq6.8}
\end{equation}
where
\begin{equation}
\dot C = \frac{C^{\rm new} }{ t_S} \, . \label{eq6.9}
\end{equation}
Here, $C^{\rm new}$ is the \emph{present} value of the capture rate of the new
population (that discussed in the previous section which assumed an overdensity
$\delta_E$ built up over the full time $t_S$) and $t_S = 4.5$~Gyr is the build
up time, i.e.\ the lifetime of the Solar System.
\pagebreak[3]

Note that the problem of determining the total, present annihilation rate, in
presence of the new population, is a non-linear one. The answer cannot be
simply split as the effect of the old population, plus the effect of the new
one. The differential equation to be solved, eq.~(\ref{eq6.2}) with
eq.~(\ref{eq6.8}), is a Riccati equation. In general, such an equation cannot
be solved analytically. However, in the present case (corresponding to the
restricted type of equations originally discussed by Riccati), the problem can
be solved in terms of Airy functions. To do that let us first scale the
variables to simplify the Riccati equation. Let us introduce the reduced
variables $x$ and $y$ by posing
\begin{equation}
C^{\rm tot} (t) = C^{\rm old} + \dot C \, t \equiv \alpha \, x \,, \qquad N
\equiv \beta \, y \, , \label{eq6.10}
\end{equation}
where
\begin{equation}
\alpha = \dot{C}^{2/3} \, C_a^{-1/3} \, , \qquad \beta = \dot{C}^{1/3} \, C_a^{-2/3}
\label{eq6.11}
\end{equation}
(the notation $\alpha$ here should not be confused with our unrelated previous
use of this letter). In terms of $x$ and $y$ the evolution equation
(\ref{eq6.2}) reads simply
\begin{equation}
\frac{dy}{dx} + y^2 = x \, . \label{eq6.12}
\end{equation}

If we now introduce a new variable $u(x)$ by posing $y(x) = (du / dx) / u$
(this use of the letter $u$ should not be confused with the notation above
for the speed with respect to the Earth),
eq.~(\ref{eq6.12}) becomes the Airy equation:
\begin{equation}
\frac{d^2 \, u}{d \, x^2} = x \, u \, . \label{eq6.13}
\end{equation}
The general solution of (\ref{eq6.13}) reads $u(x) = a \, {\rm Ai} \, (x) + b
\, {\rm Bi} \, (x)$, where ${\rm Ai} \, (x)$ is the usual Airy function
(decreasing for $x \rightarrow + \infty$) and ${\rm Bi} \, (x)$ the
complementary solution (increasing as $x \rightarrow + \infty$). The boundary
condition of our problem is that $y$ \emph{vanishes} when $x$ takes it initial
(positive) value
\begin{equation}
x_i = \frac{C^{\rm old}}{\alpha} \label{eq6.14}
\end{equation}
(corresponding to $t=0$). We are then interested in the final value $y_f = y
(x_f)$ of $y$ corresponding to the final (present) value of $x$:
\begin{equation}
x_f = \frac{C^{\rm tot} \, (t_S)}{\alpha} = \frac{C^{\rm old} + \dot C \,
t_S}{\alpha} \, . \label{eq6.15}
\end{equation}
This final value is given by
\begin{equation}
y_f = \frac{{\rm Ai}' (x_f) \, {\rm Bi}' (x_i) - {\rm Bi}' (x_f) \, {\rm Ai}'
(x_i)}{{\rm Ai} (x_f) \, {\rm Bi}' (x_i) - {\rm Bi} (x_f) \, {\rm Ai}' (x_i)}
\, . \label{eq6.16}
\end{equation}
In terms of $y_f$ the total, present annihilation rate reads
\begin{equation}
\Gamma_a^{\rm tot} (t_S) = \frac{1}{2} \ \alpha \, y_f^2 \, . \label{eq6.17}
\end{equation}

\FIGURE[t]{
\epsfig{file=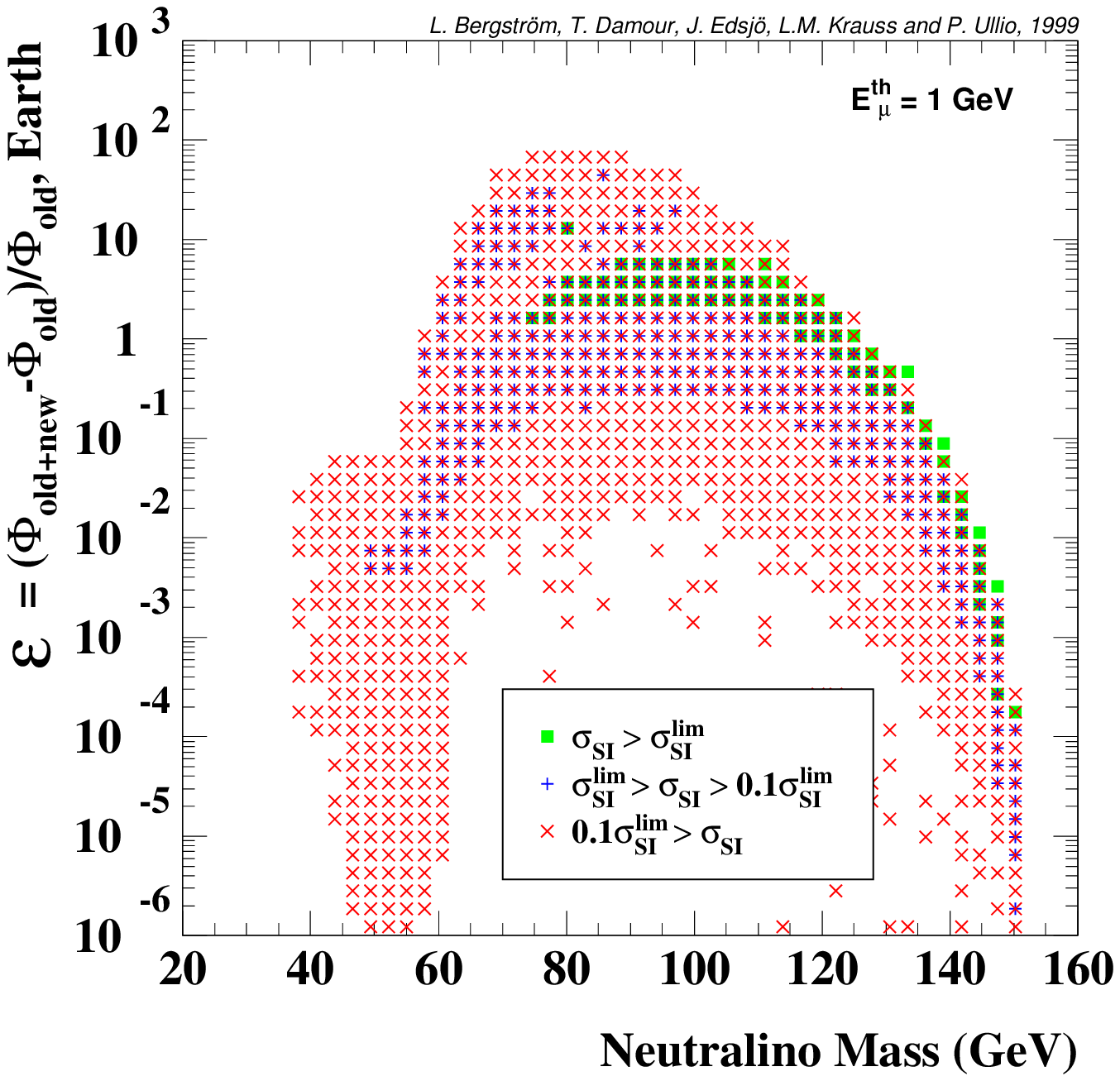,width=0.75\textwidth}
\caption{The enhancement factor, eq.~(\ref{eq:enh}), as a function of
  the neutralino mass.  The points are symbol coded according to
  whether they exceed the claimed direct detection bound $\sigma_{SI}$
  \protect\cite{uppbnd} from the {\sc Dama} experiment (squares), give
  direct detection rates within a factor of 10 of those limits
  (pluses) or give rates more than 10 times smaller than these limits
  (crosses).}
\label{fig:readknr-mx}
}

It is useful to consider also the amplification factor ${\cal A}$ brought by
the new population, i.e.\ the ratio
\begin{equation}
{\cal A} \equiv 1 +{\cal E} \equiv \frac{\Gamma_a^{\rm tot} (t_S)}{\Gamma_a^{\rm old}} = \frac{\alpha
\, y_f^2}{C^{\rm old} \, \tanh^2 \, (\gamma \, t_S)} \,,
\label{eq6.18}
\end{equation}
where we have also defined the enhancement factor ${\cal E} \equiv
{\cal A}-1$. Note that since the muon flux is directly proportional to
the annihilation rate, $\Phi_{\mu} = k\; \Gamma_a$, with $k$ depending
on the neutralino mass and the `hardness' of the annihilation
spectrum, the enhancement factor reads
\begin{equation}
    {\cal E} = \frac{\Gamma_{a}^{\rm tot} - \Gamma_{a}^{\rm old}}
    {\Gamma_{a}^{\rm old}} =
    \frac{\Phi_{\mu}^{\rm old+new}-\Phi_{\mu}^{\rm old}}
    {\Phi_{\mu}^{\rm old}}\,.
    \label{eq:enh}
\end{equation}

In fig.~\ref{fig:readknr-mx} the enhancement factor for our set of
MSSM models is shown against the neutralino mass.  As anticipated, the
\pagebreak[3]
\looseness=1 amplification mechanism is only operative for masses smaller than
around 150 GeV. Also, for small masses (less than around 60 GeV) the
enhancement is relatively small.  This is due to a combination of
factors, one being that an important contributor to the scattering in
the Sun which gives rise to the radial, ``planetary'' orbits is iron
(in fact, it is important also for the capture in the Earth), and for
WIMP masses smaller than iron the momentum transfer is less effective.
In the WIMP mass region between 60 and 130 GeV, on the other hand,
enhancements up to a factor of 100 are possible.  This can potentially
be very important for the discovery potential for dark matter of
neutrino telescopes.

The points in the figure are symbol-coded according to what direct
detection rates they correspond to. In particular, the squares
indicate neutralino models which exceed the claimed {\sc Dama}
experiment upper limit~\cite{uppbnd}, assuming a halo WIMP density of
0.3 GeV/cm$^3$. In computing the limits from {\sc Dama} we have
included the effects of the new population, but because of the low
velocities their effect on present-day detectors with their relatively
high recoil energy thresholds is quite small, and in fact hardly
visible in our figures.

\section{Muon fluxes}

\FIGURE[t]{%
\epsfig{file=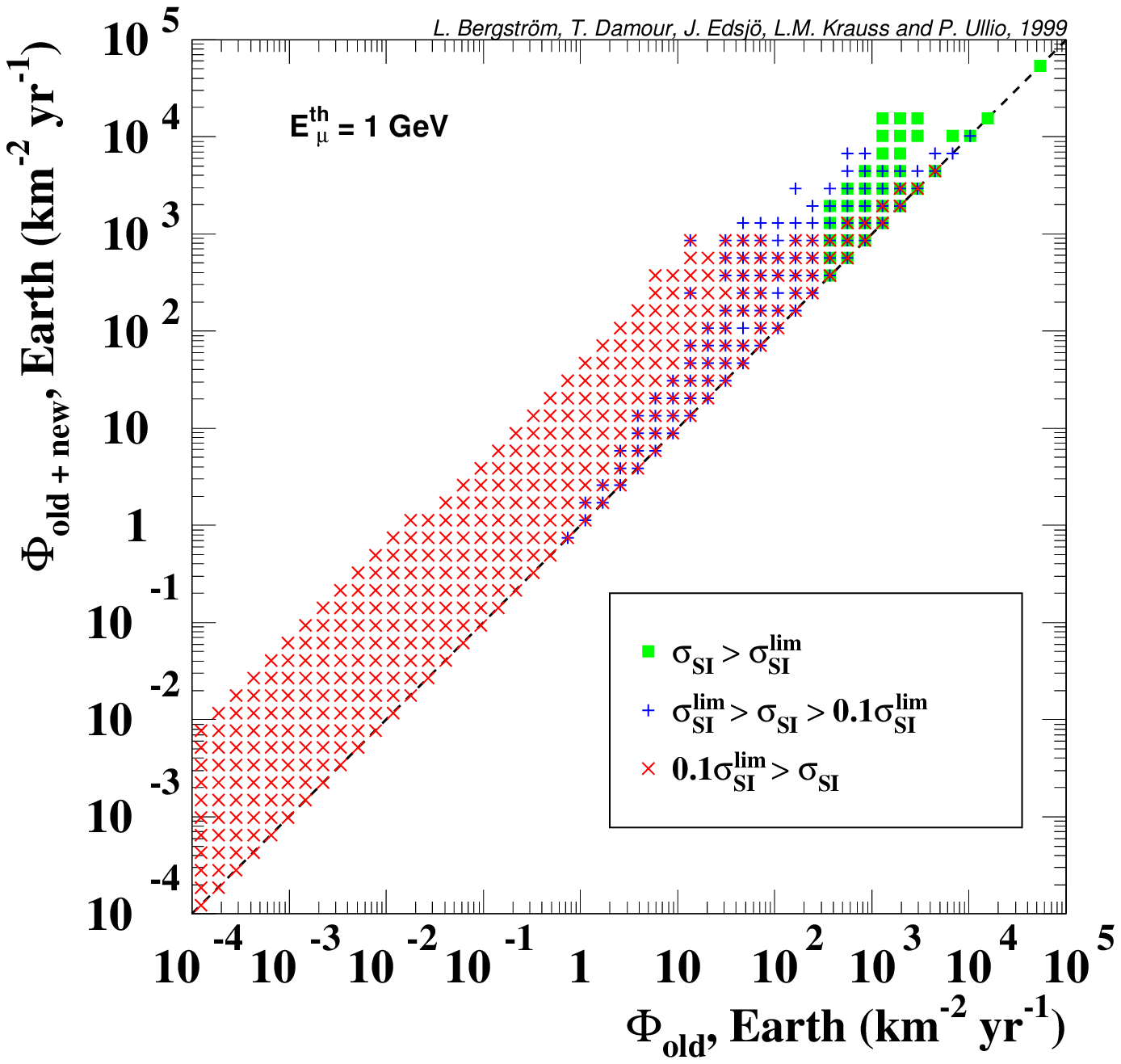,width=0.75\textwidth}
\caption{The muon flux including both the old and the new population
versus the flux obtained if only the old contribution is included.
The points are symbol coded according to whether they
exceed the claimed direct detection bound $\sigma_{SI}$ \protect\cite{uppbnd}
from the {\sc Dama} experiment (squares),
give direct detection rates within a factor of 10 of those
limits (pluses) or give rates more than 10 times smaller than
these limits (crosses).\label{fig:readk-old}}}

We calculate the resulting neutrino-induced muon fluxes essentially as
described in~\cite{km2}.  The decay and/or hadronization of the
annihilation products as well as the neutrino interactions are
simulated with {\sc Pythia} 6.115~\cite{pythia}.  All two-body
annihilation final states (at tree level) are included as well as the
1-loop induced 2 gluon and $Z$ gluon.  We use a value of $0.3$
GeV/cm$^3$ for the local neutralino mass density (leaving a study of
the change in our analysis caused by variations in this and other
astrophysical parameters to future work).

We compare the predicted rates with the present experimental upper
bound, where we have taken the Baksan results~\cite{baksan} as a
representative (the MACRO~\cite{macro} limits are very similar). The
muon threshold is set to 1 GeV. In fig.~\ref{fig:readk-old} we show
our predicted total muon rates from the direction of the center of the
Earth, compared to the rates from the ``old'' population alone. (Note,
that due to the non-linearity of the rate equations, one cannot just
sum the contributions of the ``new'' flux and the ``old'' flux
separately.) As can be seen, enhancements between one and two orders
of magnitude are frequent (even though the enhancement for the models
with the highest predicted absolute flux is generally somewhat
smaller).

\FIGURE[t]{%
\epsfig{file=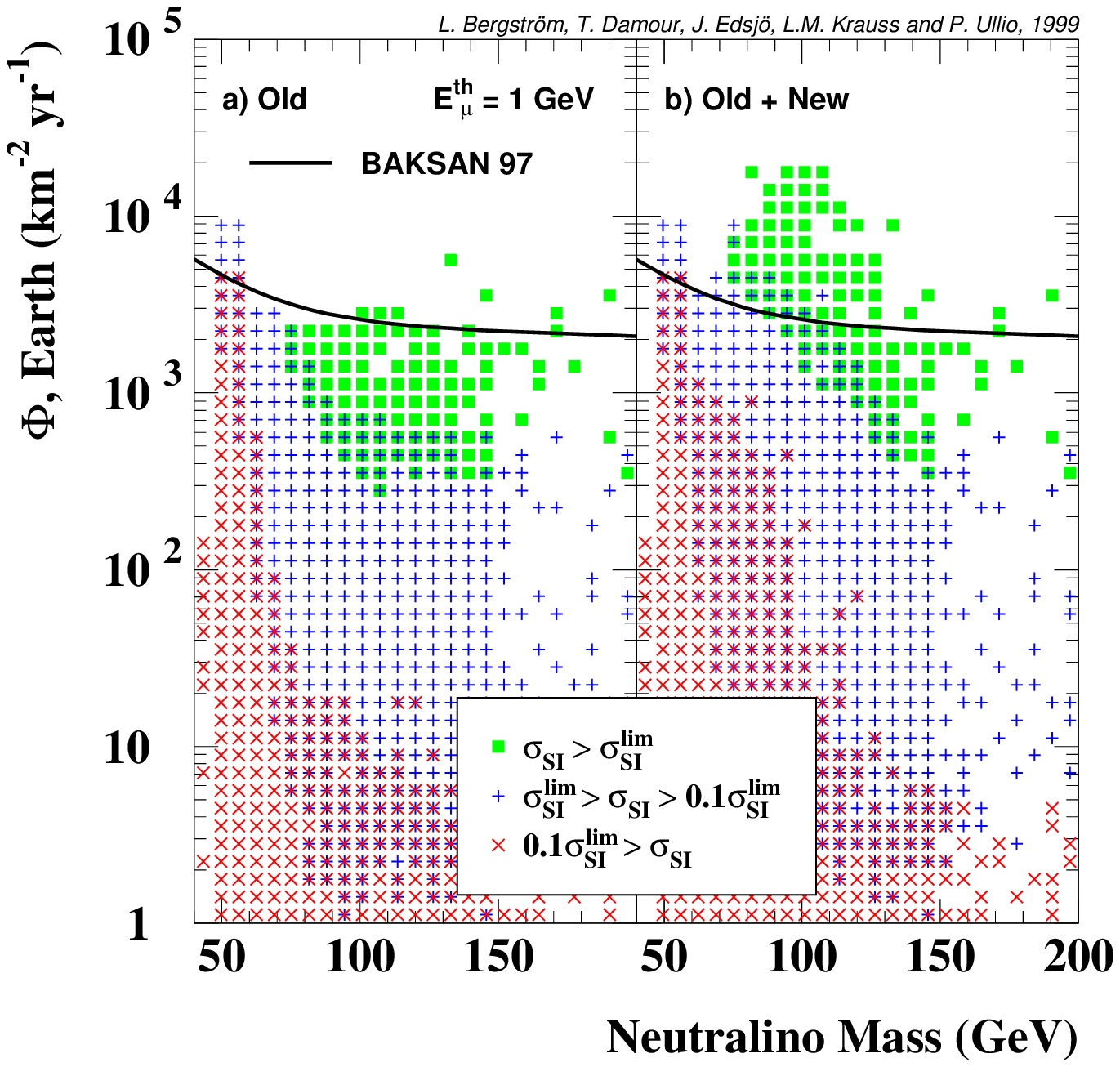,width=0.75\textwidth}
\caption{The muon flux obtained from a) only the old population and b)
both the old and the new population of WIMPs.
The points are symbol coded according to whether they
exceed the claimed direct detection bound $\sigma_{SI}$ \protect\cite{uppbnd}
from the {\sc Dama} experiment (squares),
give direct detection rates within a factor of 10 of those
limits (pluses) or give rates more than 10 times smaller than
these limits (crosses).}
\label{fig:readk-mx}}

\looseness=1 In fig.~\ref{fig:readk-mx} we show the predicted absolute fluxes
against neutralino mass for our sample of models, without and with the
new population.  The Baksan limit is shown as the nearly horizontal
line.  As can be seen, the effect on this scatter plot of the new
population is striking. Many models around 60--130 GeV have been
boosted up to higher fluxes. Several are now above the {\sc Baksan}
limit and others are within detectability in the near future. Around
80 GeV, the increase in flux for \mbox{models} with high fluxes can be as
high as a factor of 10 (for low flux models, the increase can be even
higher as seen already in fig.~\ref{fig:readk-old}).
One may note that some of the models previously thought to
be allowed by Baksan but above the claimed {\sc Dama} bound, should
really be considered as ruled out by Baksan with even greater
confidence.

\looseness=1 We have investigated the composition of the neutralinos that cause the
largest enhancement of the muon flux and that at the same time have
high absolute fluxes. We find that they are generally of
mixed type (neither very pure Higgsino nor gaugino), and that the biggest
enhancements occur for neutralinos which simultaneously have large
spin-dependent and spin-independent cross sections. We interpret this
as being due to the fact that a large spin-dependent cross section implies
efficient scattering on hydrogen in the Sun, whereas a large spin-independent
cross section is necessary for capture in the Earth (which is composed
of spinless nuclei). The highest fluxes also occur for rather light
masses of the CP-odd Higgs boson, $m_{H_3^0} \lsim 150$~GeV\@.

In future experiments, it may be realistic to reach a sensitivity of
the order of a few events per km$^2$ per year, covering most of the
area in the flux versus mass diagram in fig.~\ref{fig:readk-mx}.  It
is obvious that the enhancements found in this work may be crucial if,
for instance, the uncorrected rate would fall just below this
value. (It is to be reminded that a priori, we have no knowledge
whatsoever about which model, if any, would represent ``the world'' in
our scans.)

The correlation between signals from the Earth and the Sun have also
been investigated. Even without this new population, there are models
for which the flux from the Earth is higher than that from the Sun,
but including this new population, we find many more models (at
60--130 GeV) for which it is more advantageous to look at the Earth
than the Sun.

We have also made runs with the angular momentum parameter $\lambda=2$.
The results are very similar, and we do not display them here. The main
effect is a shift of the kinematical cut-off at 150 GeV to around 170
GeV, and an increased muon rate by up to 50 \% in some cases.

\section{Conclusions}

We have investigated the contribution from a new population of WIMPs
coming from WIMPs that have scattered in the outskirts of the Sun and
(due to small perturbations by the other planets) are confined to
nearly radial orbits reaching out to the Earth.

We find that even for the conservative case when $\lambda=1$ (i.e.\
the $z$ component of the angular momentum is conserved), we can get
enhancements of up to a factor of 100 when the neutralino mass is less
than around 150 GeV\@.

For the mass range of 60--130~GeV some models can already be ruled out
on the basis of existing data. The next generation of
detectors should be able to probe a much larger region of parameter
space in this mass range than would otherwise be possible.  

As a result of this new WIMP population, for certain
models with WIMPs in the mass range 60--130~GeV, there can be an even 
greater advantage than before in looking for the indirect signal from the
Earth rather than from the Sun.

Whether enhanced annihilations in the Earth for larger WIMP masses
might be possible depends on knowing the precise details of the Solar
System WIMP velocity distribution, which will await further numerical
work on the evolution of Solar System WIMPs.

\acknowledgments

L.B.\ was supported by the Swedish Natural Science Research Council
(NFR).  T.D.\ was partially supported by the NASA grant
NAS8-39225 to Gravity Probe B (Stanford University).
The research of L.M.K. was supported in part by the U.S.
Department of Energy.


\end{document}